\documentclass[fleqn,twoside]{article}
\usepackage{espcrc2}

\usepackage{graphicx}
\usepackage{epsfig}

\newcommand{\AmS}{{\protect\the\textfont2
  A\kern-.1667em\lower.5ex\hbox{M}\kern-.125emS}}
\newcommand{\alphas}{\mbox{$\alpha_{\mathrm{s}}$}}
\newcommand{\QQ}{\mbox{$Q^2$}}

\newcommand{\ET}{\mbox{$E_T$}}
\newcommand{\eTj}{\mbox{$E_T^{\,\mathrm{jet}}$}}

\title{Inclusive Jets and \alphas\ at HERA}

\author{P. J. Bussey\address
       {Department of Physics and Astronomy, University of Glasgow, 
        Glasgow G12 8QQ, United Kingdom\protect\\[3mm]
        for the H1 and ZEUS Collaborations}}
       
\begin{document}

\begin{abstract}
A survey is given of recent HERA results in jet production  
and prompt photon production, together with the evaluation of
the QCD coupling constant \alphas\ by a variety of techniques. 
\vspace{1pc}
\end{abstract}

\maketitle

\section{INTRODUCTION}
The presence of quarks and gluons in the proton, together with the
varied behaviour of the photon, mean that $ep$ collisions at the HERA
collider generate a wide-ranging variety of processes in which QCD can
be studied and tested.  Since 1996, HERA has accumulated well over 100
pb$^{-1}$ of data in each of the major collider experiments, running
with 27.5 GeV electrons and positrons, and protons at 820-900 GeV.  In
this review, a number of recent results connected with jet production
are discussed.  The primary concern is with the hard QCD scatter that
generates the jets, but the properties of a final-state jet itself are
also able to provide interesting perspectives on QCD.  In both cases,
it is now found to be necessary to employ a next-to-leading order
(NLO) treatment in perturbative calculations.  An important objective
is to obtain consistent and accurate measurements of the QCD coupling
constant \alphas.

\section{INCLUSIVE JETS IN PHOTOPRODUCTION AND DIS}

\begin{figure}[b!]
\hspace*{3ex}\epsfig{file=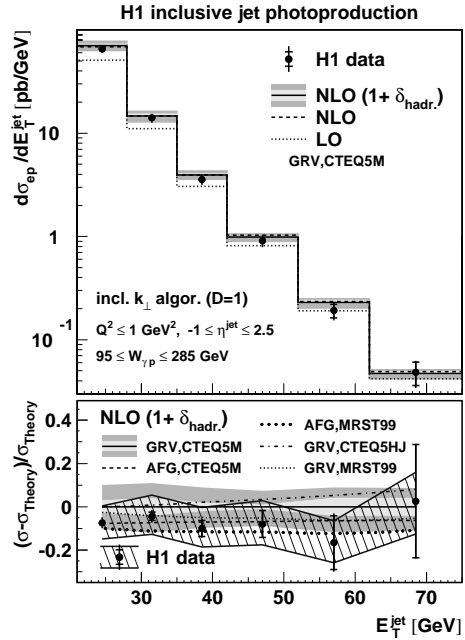,width=6cm}\\[-7ex]
\caption{Transverse energy in inclusive jet photoproduction in H1.
Data are compared with NLO predictions (Frixione and Ridolfi) using
various photon and proton PDFs, with uncertainty bands as indicated.}
\end{figure}
Photoproduction at HERA implies low values of the virtuality
\QQ\ of the virtual photon exchanged in the $ep$ collision.  The lowest-order
diagrams in \alphas\ here correspond to 2$\to$2 processes involving
the incident photon itself (direct processes) or a quark or gluon
within the photon (resolved processes).  The scatter off a quark or
gluon in the proton then gives outgoing high-\ET\ jets.  Such
processes depend on \alphas, and a well-defined measurement of the cross
section can enable \alphas\ to be determined.

It is different with the most elementary process in deep inelastic
scattering (DIS), where the incident virtual photon is simply absorbed
by a quark in the proton, ejecting it to generate a single observed
jet in the laboratory.  This process has no \alphas\ dependence, and
is sensitive merely to the parton density functions (PDFs) in the
proton and to the electromagnetic coupling of the quarks.
\begin{figure}[t]\vspace*{-5.9ex}
\hspace*{-3ex}
\epsfig{file=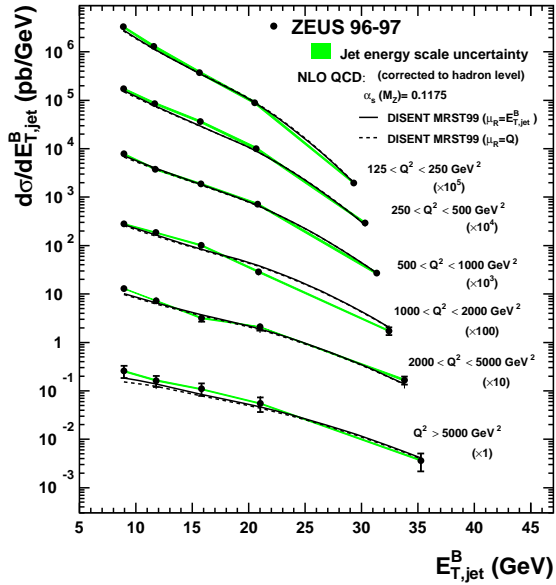,width=9cm}\\[-7ex] 
\caption{Inclusive jet cross sections in the Breit frame in ZEUS,
for different ranges of \QQ. Scaling violations are evident. 
Comparison is made to NLO calculations using DISENT. 
}
\end{figure}

To maximise sensitivity to \alphas\ in DIS, a good procedure is
to measure jets in the Breit frame.  Here, the lowest-order
jet is found at low \ET.  A selection on  high-\ET\ jets thus requires
at least two jets to be produced, and so suppresses the \alphas$^0$
contribution to the inclusive jet cross section.  By a suitable 
choice of angular cuts, the effects of the proton remnant can also
be suppressed, and sensitivity to \alphas\ can be attained.

In all cases, it is necessary to be convinced that a good description
of the process is being obtained within QCD. Next-to-leading-order
(NLO) calculations are now widely available, and the parton-level
cross sections can be corrected to hadron-level cross sections by
means of a program such as ARIADNE which contains a well-tried
description of the fragmentation within an adequate approximation to
NLO QCD.  Figs.\ 1 and 2 illustrate inclusive jet production cross
sections from H1 and ZEUS, in photoproduction and DIS (Breit frame)
respectively. A lowest-order (LO) calculation is inadequate. Within
the present experimental and theoretical uncertainties, it is
concluded that the use of NLO QCD is both necessary and sufficient to
account for the data.  There is little sensitivity to the proton and
photon PDFs provided that recent models are used.

Fig.\ 3 illustrates a recent analysis of inclusive jet cross sections
in photoproduction from ZEUS in terms of a scaling
variable. Again, the importance of an NLO approach is clearly seen,
the LO calculation being far too low. By showing that the scaled
jet invariant cross sections vary with the $\gamma p$ centre-of-mass
energy, scaling violations in photoproduction were for the first time
demonstrated in this study.

\begin{figure}[t]
\vspace*{-2ex}
\epsfig{file=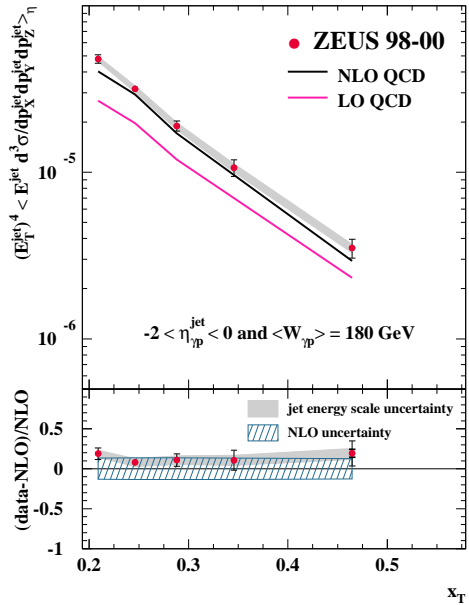,width=8.5cm}\\[-7ex]
\caption{Photoproduced scaled jet cross sections in ZEUS,  
as a function of the scaling variable $x_T=2\eTj/W_{\gamma p}$,
compared with LO and NLO QCD calculations.}
\end{figure}

\begin{figure}[t]
\hspace*{5ex}\epsfig{file=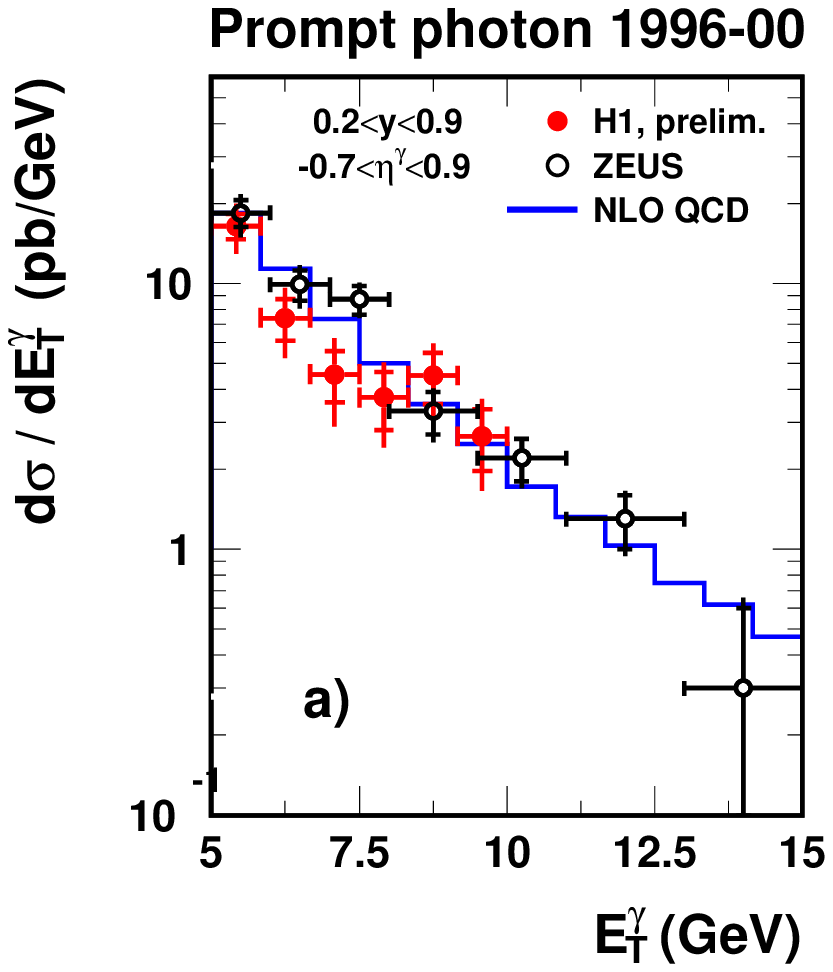,width=5cm}\\[-3ex]
\hspace*{5ex}\epsfig{file=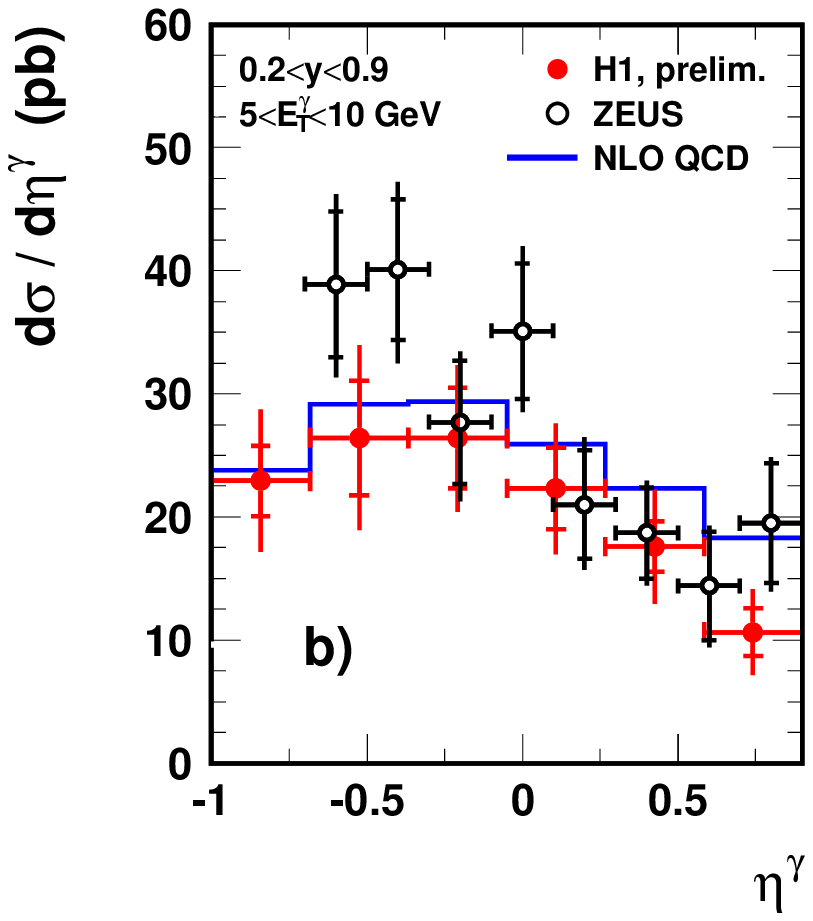,width=5cm}
\\[-6ex]
\caption{Inclusive prompt photon cross sections from H1 and ZEUS, 
compared with NLO prediction,
as a function of (a) transverse photon energy (b) pseudorapidity.}
\end{figure}
\section{PROMPT PHOTONS AT HERA.}

In a different type of QCD process, a photon may be produced in
partonic scatters instead of a quark or gluon.  Measurements of these
photons provide a further perspective on QCD of interest because the
photon emerges directly from the QCD scatter without the fragmentation
which effects the conversion of a quark or gluon into an observable
jet.  H1 have presented new measurements of inclusive prompt photons in
photoproduction, which they have compared with earlier ZEUS
measurements and with an NLO theoretical prediction from Fontannaz, Guillet
and Heinrich (using AFG and MRST2 PDFs for the photon and proton
respectively).  The results are illustrated in Fig.\ 4.  Bearing in
mind the different calorimetric techniques used by the two experiments
to identify the photons, the results are in reasonable agreement with
each other and with the theory.  However more statistics are needed to test
the theory in more depth.

ZEUS have given the first measurements of prompt photon production in
DIS.  The results, show a fair agreement with PYTHIA and HERWIG and
(Fig.\ 5) with a new NLO calculation by Kramer and Spiesberger.
The systematic errors are large, however, and an increase in
statistics would be highly desirable.

\begin{figure}
\vspace*{-1ex}
\hspace*{10ex}\epsfig{file=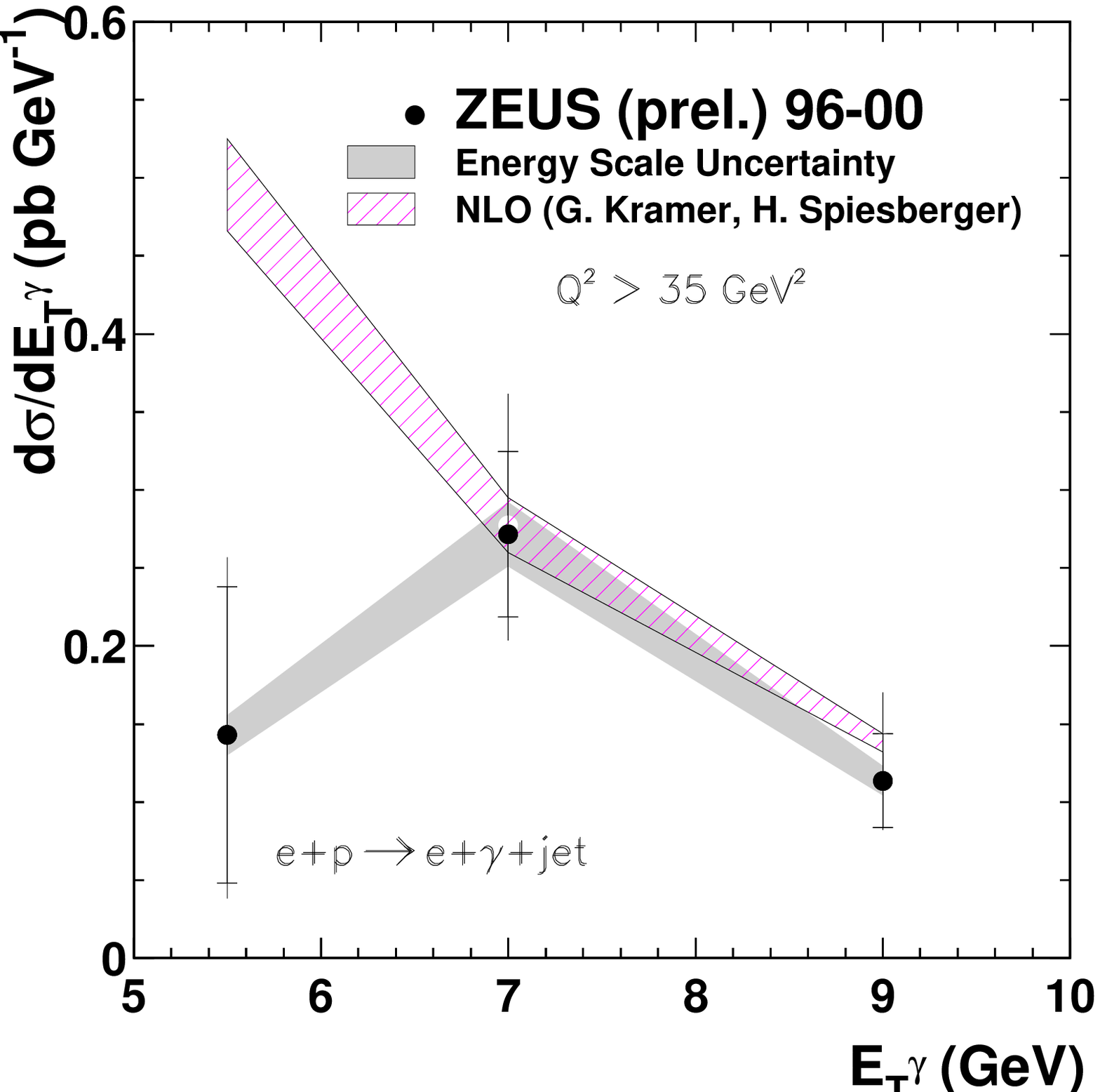,width=5cm}\\
\hspace*{10ex}\epsfig{file=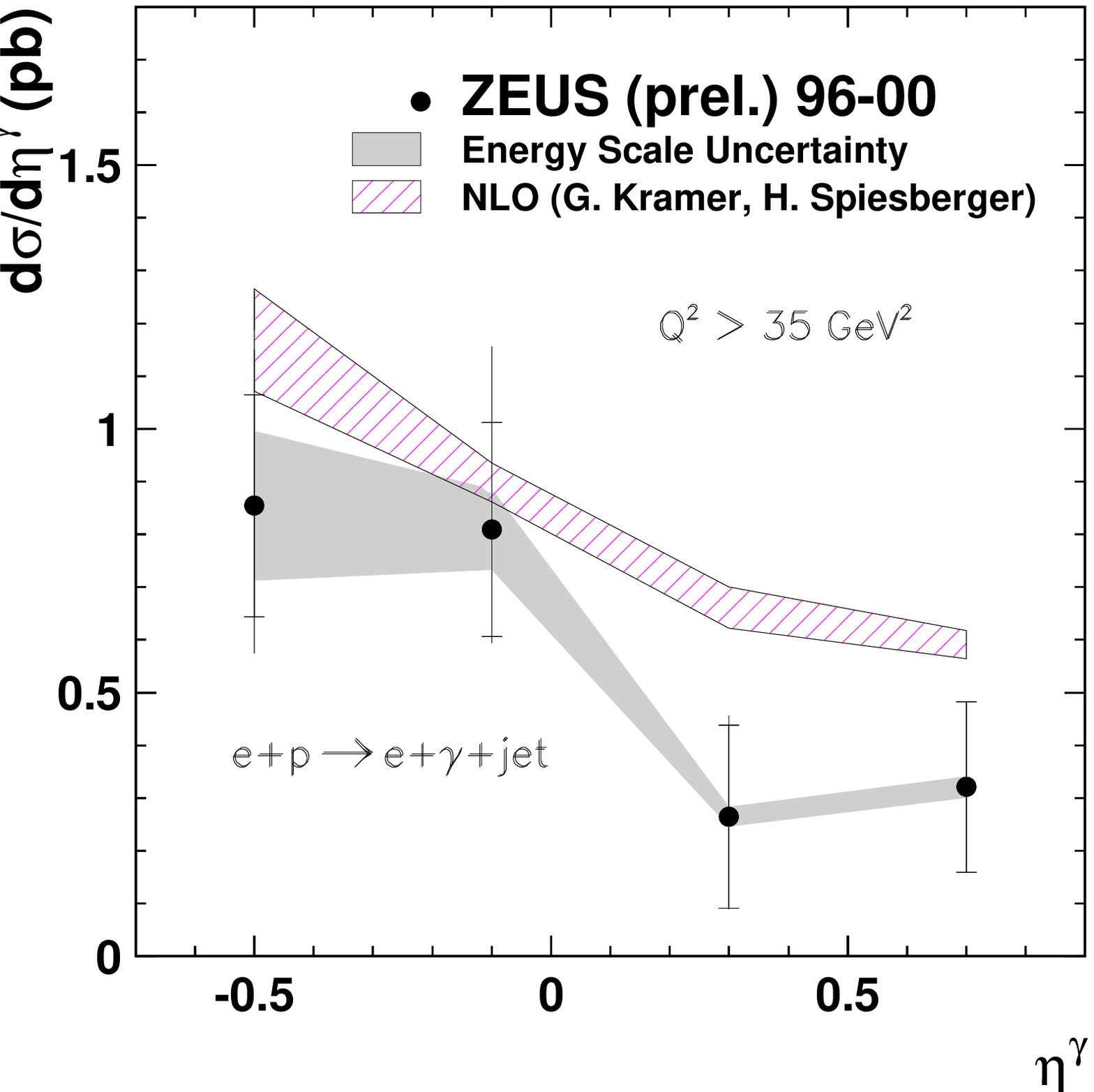,width=5cm}
\\[-4ex]
\caption{Prompt photon cross sections in DIS from ZEUS,
accompanied by a jet, compared with NLO predictions.}
\end{figure}

\section{MULTIJET FINAL STATES}

A further means of testing QCD at HERA comes from the study of final
states containing more than two jets.  Such states can arise in
several ways in photoproduction.  Since the photon fluctuates into a
meson-like intermediate state, there is a possibility for multiparton
interactions (MPI), i.e.\ more than one parton-parton scatter occurs
within the context of a given photon-proton interaction. Alternatively
expressed, there may be a soft underlying event (SUE) in addition to a
given harder scatter. Hard initial or final state gluon radiation may
take place, which is modelled in certain basically LO calculations
such as PYTHIA and HERWIG.  More generally, the processes should be
calculable in a suitably higher order of QCD.

\begin{figure}
\vspace*{0.7ex}
\hspace*{8ex}\epsfig{file=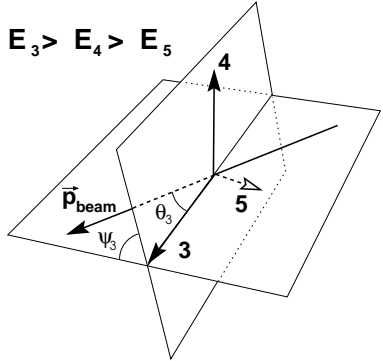,width=5cm}
\\[-4ex]
\caption{Definitions of geometric angles in three-jet final 
states (of energy $E_3$, $E_4$, $E_5$) in HERA processes.}
\end{figure}

In an $ep$ process that gives rise to three final-state partons or
jets, $1+2 \to 3+4+5$, the geometry of the final state is usefully
described by means of angles as shown in Fig.\ 6.  ZEUS have made a
preliminary study of four-jet final states, merging together the two
jets with smallest combined invariant mass so as to be able to use
this geometrical description.  Jets of $\eTj > 5$ GeV are used.  The
$\theta_3$ distribution is found (fig.\ 7) to be poorly described by
HERWIG without or with its standard SUE option, but satisfactorily
when an eikonal model (`Jimmy') is employed to generate a second
parton scatter. PYTHIA with its MPI option gives a satisfactory
result.

\begin{figure}[t!]
%\vspace*{-4.5ex}
\hspace*{5ex}\epsfig{file=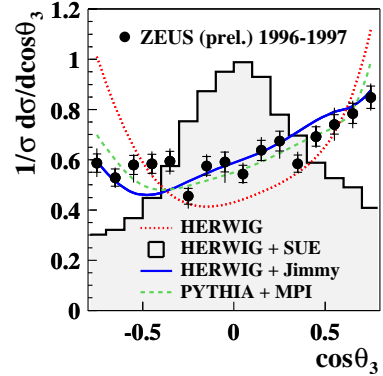,width=5cm,%
bbllx=24pt,bblly=270pt,bburx=281pt,bbury=520pt,clip=yes}
\\[8ex]
\hspace*{5ex}\epsfig{file=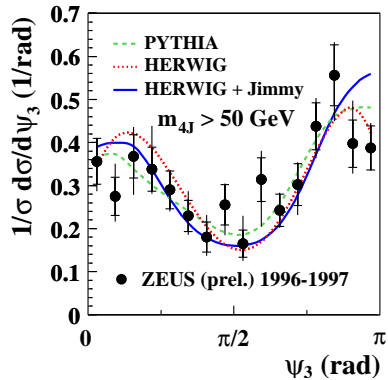,width=5cm,%
bbllx=24pt,bblly=264pt,bburx=281pt,bbury=520pt,clip=yes}
\\[-3ex]
\caption{Normalised distribution in the angle $\theta_3$
measured by ZEUS, compared with models including multiparton 
effects, and with a mass requirement on the four-jet final state.
}
\end{figure}

Since the SUE/MPI processes interfere with QCD studies, it is
desirable to eliminate them by suitable requirements on the event.  If
it is insisted that the combined mass of the four jets exceeds 50 GeV,
both HERWIG and PYTHIA give good descriptions of the distribution, and
the SUE/MPI effects are much reduced.  This condition applies to the
kinematics of the events used in the inclusive jet measurements
discussed above.

In DIS, the SUE/MPI effects should be absent or largely suppressed. H1 have
compared three-jet distributions with the NLOJET calculation (Nagy
and Tr\'ocs\'anyi) evaluated at LO and NLO.  Results are illustrated in
Fig.\ 8. While both versions describe well the shape of the angular 
distributions, and phase space does not, only the NLO calculation
correctly predicts the cross section as a function of
\QQ.

\begin{figure}[t]
\vspace*{-0.5ex}
\hspace*{4ex}\epsfig{file=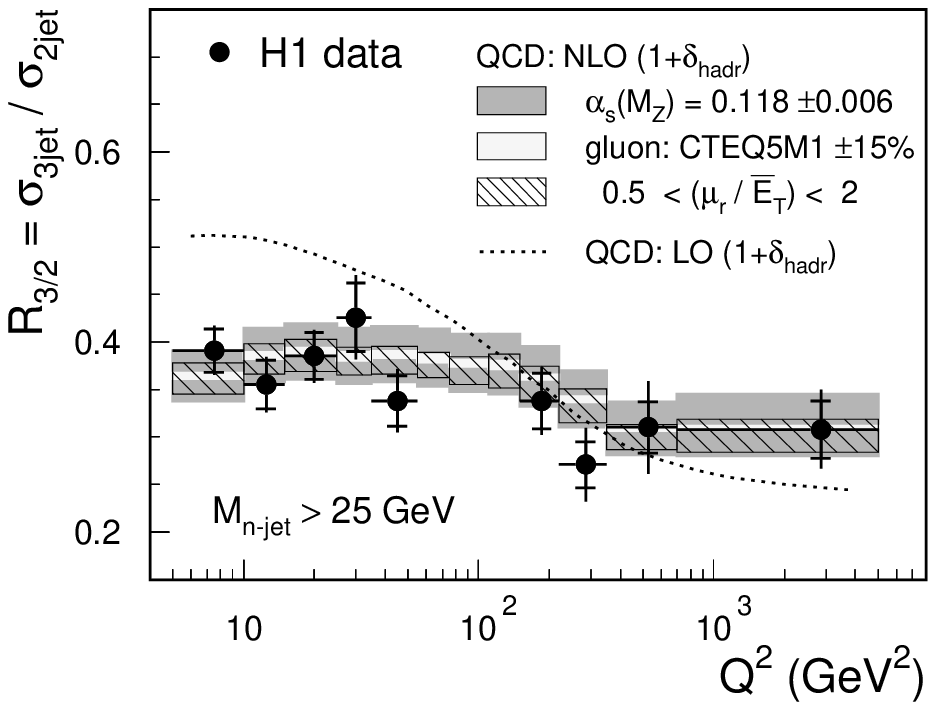,width=6cm}\\[3ex]
\mbox{\hspace*{0ex}\epsfig{file=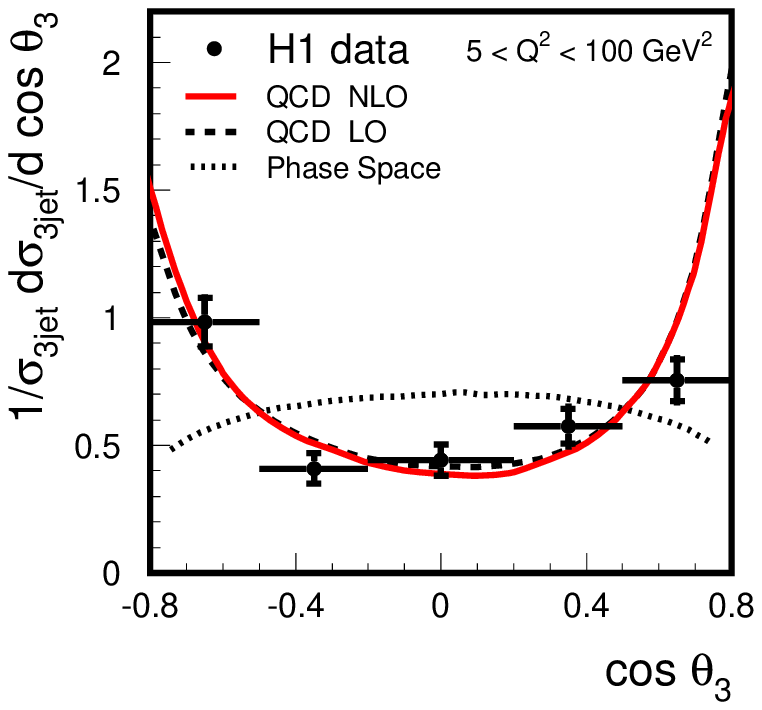,width=3.7cm}
\hspace*{0ex}\epsfig{file=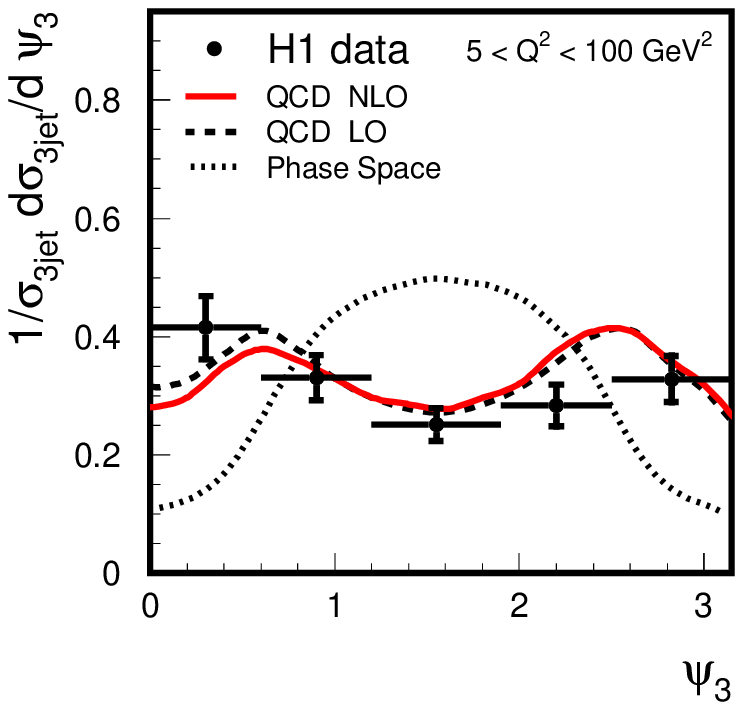,width=3.7cm}}
\\[-4ex]
\caption{(Upper) 3-jet/2-jet ratio, (lower)  $\theta_3$ and $\psi_3$ 
distributions in DIS
measured by H1, compared with LO and NLO
predictions and with phase space.  }
\end{figure}

\section{DETERMINATION OF \alphas}

The analyses presented above have shown in a variety of ways that NLO
calculations are able to give a good description of QCD-governed
processes at HERA involving outgoing jets.  It is therefore possible
to use these processes with assurance to evaluate the QCD coupling
constant \alphas, even though the experimental situation is
intrinsically less clean than in $e^+e^-$ colliders.  A number of
accurate \alphas\ measurements have been published by H1 and ZEUS in
recent years.  One technique, in which no explicit jet measurement is
required, is to carry out global QCD-based fits to the structure
functions measured in DIS, in the course of which, \alphas\ is
determined.  The results of these determinations are tabulated below.

ZEUS have used the specific measurement of inclusive jets in both DIS
and photoproduction to measure \alphas.  In the Breit Frame
measurements described above, an accurate determination of \alphas\
requires that an explicit dependence of the proton PDFs on \alphas\ in
an NLO context be taken into account, which can be done in the MRST99
parameterisation.  In their analysis of photoproduced jets, ZEUS 
fit the measured jet cross sections 
to the NLO calculation of Klasen,
Klainwort and Kramer. Figs.\ 3 and
9 illustrate the effects of the theoretical and experimental
uncertainties on this measurement.

\begin{figure}[t!]
\vspace*{-7.5ex}
\hspace*{-4ex}
\epsfig{file=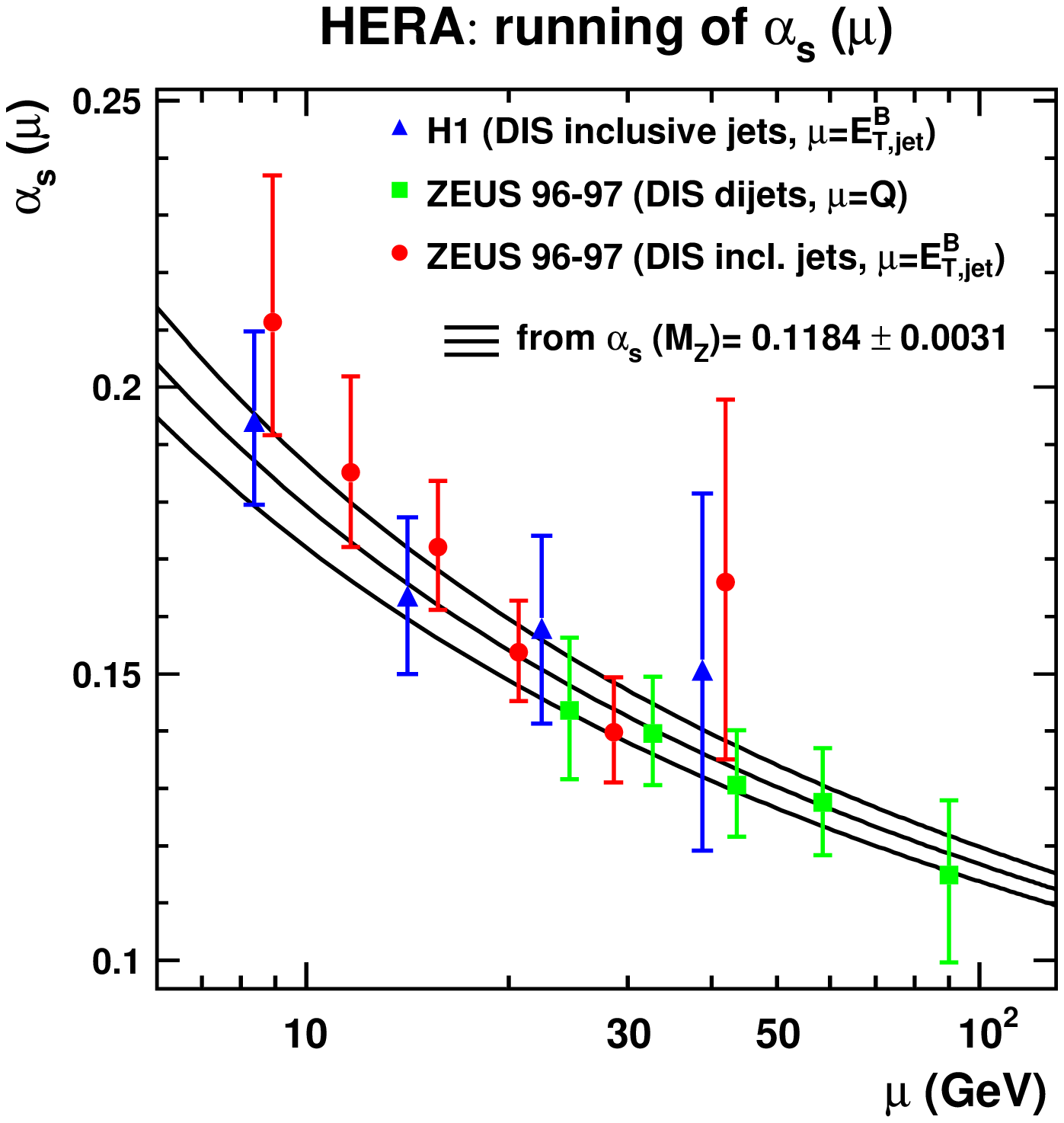,width=9cm}\\[-15ex]
\hspace*{-4ex}
\epsfig{file=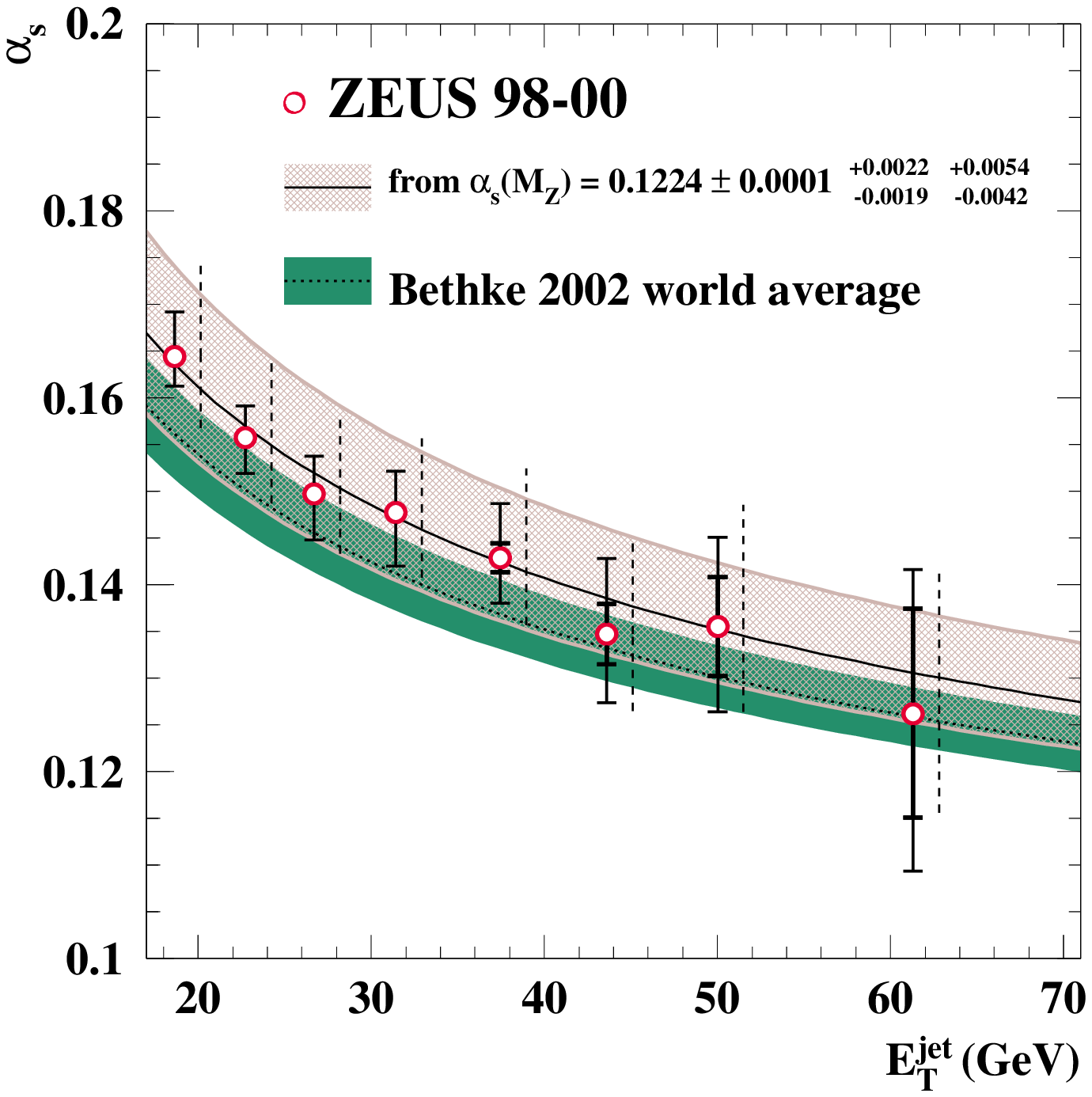,width=9cm}\\[-11ex]
\caption{Fitted values of \alphas\ obtained in various HERA
analyses, plotted as a function of the QCD scale variable.  Upper
plot: scale variable ($\mu$) taken as the dominant momentum transfer
in the process, namely the unsquared photon virtuality or the
transverse jet energy. Lower plot: taken as the transverse energy of
jets in photoproduction.  In both cases comparison is made with a
recent world average value of \alphas\ shown with its
uncertainty.\protect\\[1ex] }
\end{figure}

In the upper plot of Fig. 9, values of \alphas\
obtained by several techniques at HERA are plotted as a function of
the QCD scale of the relevant hard interaction, as indicated in the
figure.  The running of \alphas\ is clearly exhibited, and the
different methods are consistent.  These conclusions are confirmed in
the lower plot of Fig.\ 9, which illustrates the consistency of the 
photoproduced jet behaviour in the ZEUS data. 

In a different approach, ZEUS have measured \alphas\ using the
structure of jets in DIS measured in the laboratory frame.  These are
predominantly quark jets.  Here, as in the two other determinations
ust discussed, a $k_T$-clustering jet finder was employed.  In such a
jet finder, there is a parameter $y_{\mathrm{cut}}$ which designates
the degree of `closeness' of particles or particle clusters which are
to be aggregated within a jet.  Normally this parameter is set to
unity, but if it is decreased, more jet-like objects become
distinguished, referred to as `subjets'. The increase of the average
total number of subjets $<\!n_{\mathrm{sbj}}\!>$ with decreasing
$y_{\mathrm{cut}}$ is sensitive to \alphas, since the first feature of
the jet substructure to be encountered is the possible final-state
radiation of a hard gluon within the overall jet.  Eventually a
sensitivity to details of the fragmentation emerges at small
$y_{\mathrm{cut}}$ values.

\begin{figure}[t!]
\vspace*{-3ex}
\epsfig{file=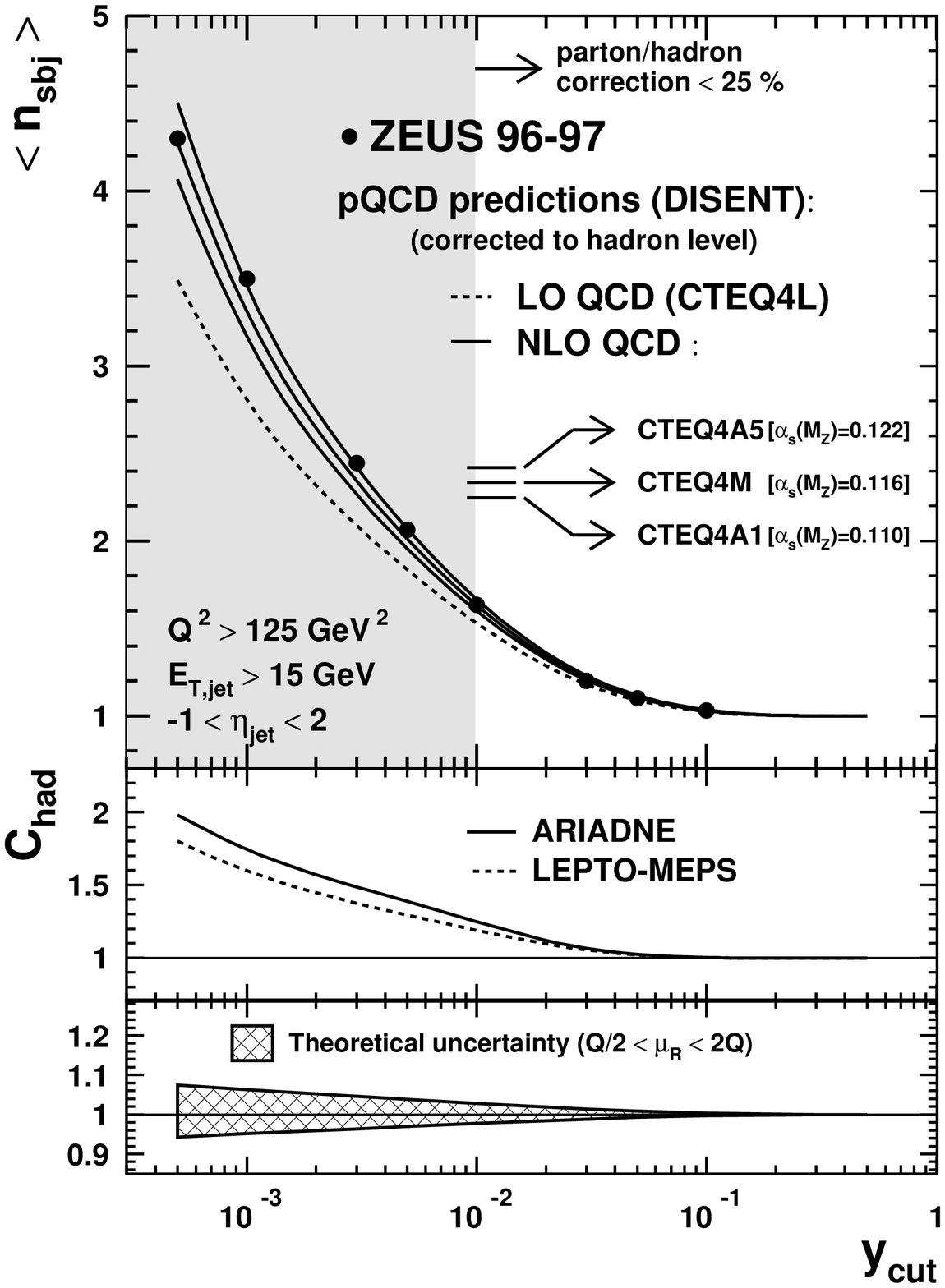,width=7cm}\\[-3ex]
\epsfig{file=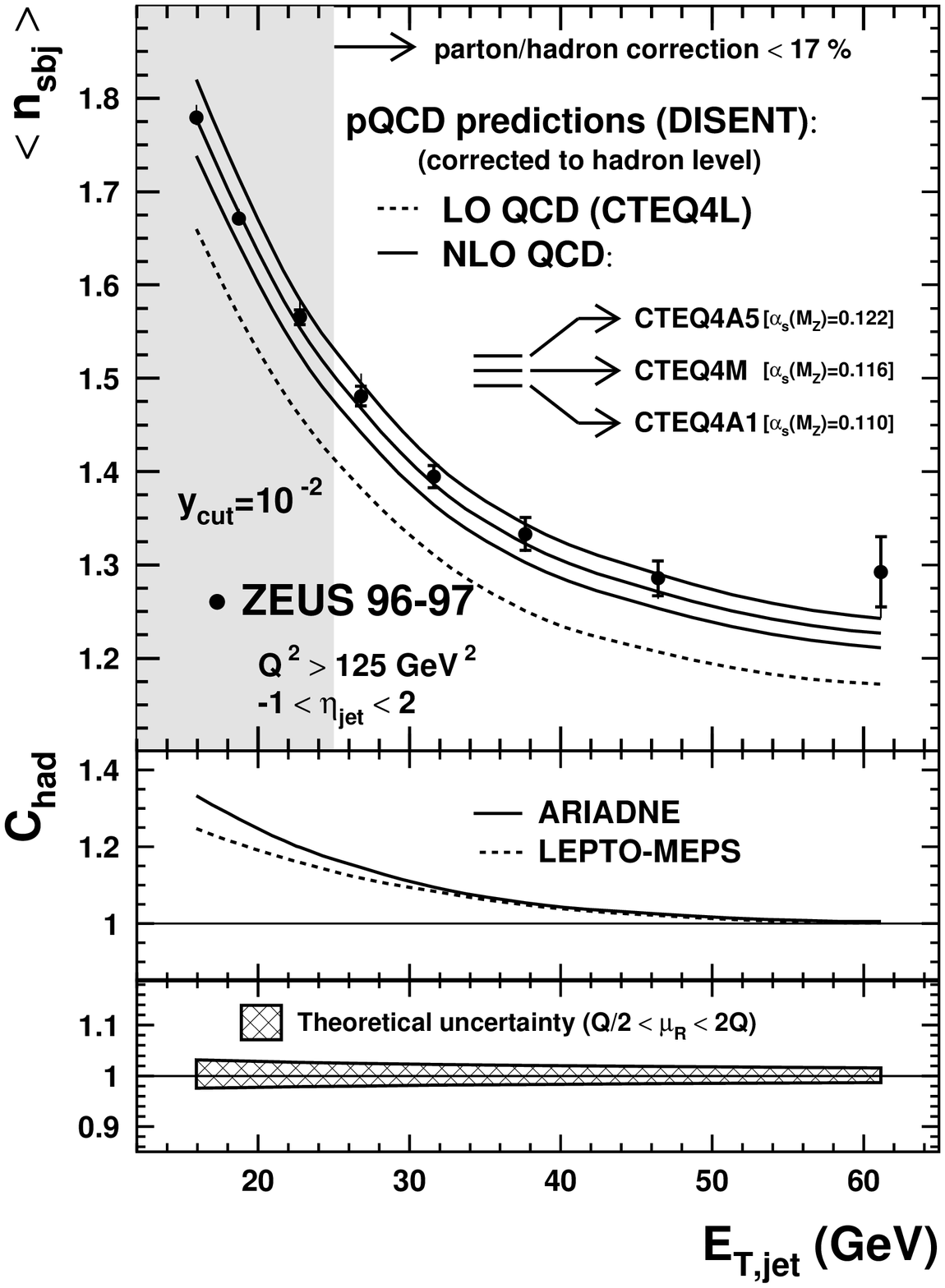,width=7cm}\\[-7ex]
\caption{Subjet multiplicity measured by ZEUS as a function of $y_{\mathrm{cut}}$,
corrected to hadron level by means of ARIADNE.  There is good
agreement with NLO QCD, but not LO. Variations of \alphas\ and proton
PDF are indicated.}
\end{figure}

By careful study it was found that the NLO parton generator DISENT,
corrected to the hadron level, was able to give a consistently
accurate description of the variation of $<\!n_{\mathrm{sbj}}\!>$ with
$y_{\mathrm{cut}}$ (Fig.\ 10. top)).  The data are insensitive to details
of the proton PDF used. Varying \alphas, a fit was
performed to $<\!n_{\mathrm{sbj}}\!>$ plotted as a function of \eTj, using a
value of $y_{\mathrm{cut}} = 0.01$ which minimises the theoretical and
experimental uncertainties (Fig.\ 10, bottom).  An overall optimal
value of \alphas\ was extracted.

\begin{figure}[b!]
\vspace*{-5ex}
\epsfig{file=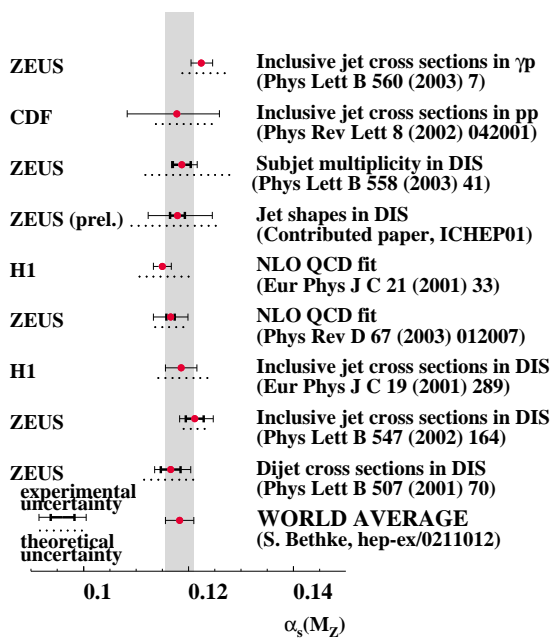,width=9.3cm}
\\[-5ex]
\caption{Summary of recent HERA and CDF \alphas\ determinations.}
\vfill
\end{figure}

The \alphas\ values obtained by these three methods are as follows, the
quoted errors being respectively the statistical, experimental systematical
and theoretical uncertainties.\\[1.5ex] Inclusive jets in DIS Breit Frame:
$${0.1212\;\pm0.0017\;^{+0.0003}_{-0.0001}\;^{+0.0028}_{-0.0027}}$$
Inclusive jets in photoproduction:
$${0.1224\;\pm0.0001\;^{+0.0022}_{-0.0019}\;^{+0.0054}_{-0.0042}}$$
DIS subjet multiplicities:
$${0.1187\;\pm0.0017\;^{+0.0024}_{-0.0009}\;^{+0.0093}_{-0.0076}}$$
These results are compared with those from the DIS fits and some other recent
measurements in Fig.\ 11.

The statistical accuracy is now reduced below
the uncertainties due to the experimental systematics and the 
NLO theoretical modelling that is needed to extract the results. 
The current world average value is quoted as $0.1183\pm0.0027$,
dominated largely by results from $e^+e^-$ colliders.
It may be concluded that the HERA measurements are very competitive
with most others, and that a consistent account of QCD is being
presented over a wide range of parton phenomena.

\section{CONCLUSIONS}

The results summarised above show the outstanding success that HERA
has achieved in studying QCD-governed processes in $ep$-initiated
reactions.  Many of the processes are investigated uniquely at HERA,
and include the generation of high-energy jets and photons from
interactions that are initiated by photons over a wide range of
virtuality.  At NLO, QCD is able to give a reliable description of
essentially all the processes studied, both in photoproduction and in
DIS.  To achieve this, recent PDFs for the proton and photon are
required, and the QCD coupling \alphas\ must be appropriately set by
means of fits to the data.  Experimental values for \alphas\ are
obtained in this way whose quality compares very favorably  with those obtained
by other means.\\[5cm]

\end{document}